\documentclass{article}
\usepackage{emulateapj}
                                                                                
\begin{document}

\title{The Birthplace of Low-Mass X-ray Binaries: Field Versus Globular
Cluster Populations}

\author{Jimmy A. Irwin}
\affil{Department of Astronomy, University of Michigan,
Dennison Building, Ann Arbor, MI 48109-1042; jairwin@umich.edu}

\begin{abstract}
Recent {\it Chandra} studies of low-mass X-ray binaries (LMXBs) within
early-type galaxies have found that LMXBs are commonly located within globular
clusters of the galaxies. However, whether all LMXBs
are formed within globular clusters has remained an open
question. If all LMXBs formed within globular clusters, the summed X-ray
luminosity of the LMXBs in a galaxy should be directly proportional to the
number of globular clusters in the galaxy regardless of where the LMXBs
currently reside. We have compared these two quantities over the same angular
area for a sample of 12 elliptical and S0 galaxies observed with {\it Chandra}
and found that the correlation between the two quantities is weaker than
expected if all LMXBs formed within globular clusters. This indicates that a
significant number of the LMXBs were formed in the field and naturally
accounts for the spread in field--to--cluster fractions of LMXBs from
galaxy to galaxy. We also find that the
``pollution" of globular cluster LMXBs into the field has been minimal within
elliptical galaxies, but there is evidence that roughly half of the LMXBs
originally in the globular clusters of S0 galaxies in our sample have
escaped into the field. This is likely due to higher globular
cluster disruption rates in S0s, resulting from stronger gravitational
shocks caused by the passage of globular clusters through the disks of
S0 galaxies that are absent in elliptical galaxies.


\end{abstract}

\section{Introduction} \label{sec:introduction}
Composed of a neutron star or black hole primary and a low-mass, typically
evolved secondary star that donates material to the primary via Roche lobe
overflow, low-mass X-ray binaries (LMXBs) are the only luminous
($L_X > 10^{36}$ ergs s$^{-1}$) class of X-ray point
source expected in significant numbers in the old
stellar populations of S0 and elliptical galaxies. It came as no surprise that
{\it Chandra}, with its subarcsecond spatial resolution, was able to resolve
dozens if not hundreds of individual LMXBs in nearby early-type galaxies,
which in some instances comprise the bulk of the X-ray emission from the galaxy.
What was surprising was the discovery that a remarkably high percentage of the
LMXBs reside within globular clusters of the host galaxies 
(e.g., Angelini, Loewenstein, \& Mushotzky 2001; Kundu, Maccarone, \& Zepf 2002;
Randall, Sarazin, \& Irwin 2004). Also unexpected was the variation in the
fraction of LMXBs within globular clusters from galaxy to galaxy, ranging from
almost 70\% in NGC~1399 to a more modest 18\% in NGC~1553 (Angelini et al.
2001; Sarazin et al.\ 2003).

In retrospect, the high fraction of LMXBs within globular clusters should not
have been surprising. In the Milky Way, globular clusters host $\sim$10\% of
the Galaxy's LMXBs, despite the fact they constitute only $\sim$0.1\% of the
Galaxy's mass. Given that early-type galaxies have a higher fraction of their
mass in globular clusters than spiral galaxies, early-type galaxies would
naturally produce more LMXBs within globular clusters than spiral galaxies
of comparable mass, even if both types of galaxies produced similar numbers of
field LMXBs.

The high stellar density within globular clusters is undoubtedly responsible
for the disproportionately high number of LMXBs within them. The increased rate
of multibody interactions within globular clusters serves to tighten existing
binaries to the point that mass transfer from the secondary onto the compact
accretor can occur much more frequently than in the field. The gravitational
capture of a potential donor star by a lone neutron star or black hole is
also much more likely within a globular cluster than in the field. Given the
much higher ($>100$ times) efficiency of creating LMXBs within globular clusters
than in the field, it is natural to ask if {\it all} (or nearly all) LMXBs are
formed within globular clusters. This was first suggested by Grindlay (1984)
for the case of Galactic X-ray bursting binaries and extended to LMXBs within
early-type galaxies by White, Sarazin, \& Kulkarni (2002).
In this scenario, LMXBs that currently
reside in the field originally formed within globular clusters but were
subsequently separated from the cluster either by a supernova kick imparted
to the system once the primary evolved, through stellar encounters within
the globular cluster, or through the tidal disruption/destruction of the
the host globular cluster.

Many X-ray properties of the field LMXBs appear indistinguishable from the
globular cluster LMXBs for the galaxies studied to date, suggesting a common
origin of the two populations. Both the X-ray luminosity functions
(Kundu et al.\ 2002; Jord\`an et al.\ 2004)
and bulk spectral properties (Maccarone,
Kundu, \& Zepf 2003; Irwin, Athey, \& Sarazin 2003; Humphrey \& Buote 2004)
of the two populations are consistent with having been drawn from the same
parent distribution. On the other hand, the variation in the fraction of
LMXBs in globular clusters from galaxy to galaxy is more easily explained
as the sum of a field population of LMXBs that scales with the mass of the
galaxy and a globular cluster population of LMXBs that scales with the number
of globular clusters, particularly if it can be confirmed
that galaxies with a larger number of globular clusters have a
higher fraction of their LMXBs within globular clusters than in galaxies with
relatively few globular clusters. Indeed, previous studies have noted this
trend in a few galaxies (Sarazin et al.\ 2003; Maccarone et al.\ 2003)
that support this notion.

One potential method of determining whether the present-day field LMXB
population originated within globular clusters or are indigenous to the field
is to investigate the spatial distribution of the field LMXBs. If they formed in
globular clusters, the LMXBs might be expected to follow the generally
flatter spatial distribution of the globular clusters (e.g., Ashman \& Zepf
1998), rather than the steeper de Vaucouleurs profile of the optical light.
Observationally, this is a
difficult effect to confirm. The limited number of field LMXBs in any given
observation does not constrain their spatial profile particularly
well. Furthermore,
hot X-ray gas present in the galaxies will raise the detection limit
of sources near the center of the galaxies, leading to an artificial flattening
of the spatial distribution of the LMXBs. In addition, it is not at all clear if
the ejected LMXBs will maintain the shape of the parent globular distribution,
since different ejection mechanisms will lead to different field LMXB
distributions. If the LMXBs are ejected from globular clusters via supernova
kicks, the spatial distribution of the field LMXBs will be more extended,
given that the expected kick velocity of several hundred km s$^{-1}$ is
approximately the same as the average space velocity of a globular cluster
within a galaxy. Conversely, simulations show that tidal disruption/destruction
of globular clusters is more efficient in the central regions of galaxies
(e.g., Vesperini 2000), which would lead to a profile of stripped LMXBs
that is steeper than the remaining globular clusters.

An alternative method of addressing this issue is to compare the total number
of LMXBs to the total number of globular clusters in a galaxy for a sample
of galaxies. If all LMXBs formed within globular clusters, there should be
a linear relation between these two quantities regardless of where the LMXBs
currently reside. On the other hand, if there is a significant population of
LMXBs created in
the field, the relation between the number of LMXBs and globular clusters
should be weaker, as the field component becomes more dominant in galaxies
with fewer globular clusters.
This would also predict that the fraction of LMXBs found within globular
clusters is larger for galaxies with more globular clusters per
unit light, which could account for the measured spread in the fraction of
LMXBs within globular clusters from galaxy to galaxy.

In practice, it is more feasible to determine the total X-ray luminosity
emanating from LMXBs rather than the total number, since each observation in
general has a different limiting luminosity for source detection. In addition,
it is necessary to normalize both the total LMXB luminosity and the number
of globular clusters by the optical light of the galaxy to compare galaxies
of different physical sizes. The latter quantity is usually referred to as the
globular cluster specific frequency, and is defined as $S_N = N_{tot} \*
10^{0.4(M_V+15)}$ (Harris \& van den Bergh 1981),
where $N_{tot}$ is the total number of globular clusters in the galaxy and
$M_V$ is the absolute visual magnitude of the galaxy.
Such a comparison of $L_X/L_{opt}$ to $S_N$ for a sample of galaxies was
originally performed by White et al.\ (2002) using {\it Advanced Satellite for
Cosmology and Astrophysics (ASCA)} data, who found
a strong correlation between $L_X/L_{opt}$ and $S_N$. His best-fit relation of
$L_X/L_{opt} \propto S_N^{1.2 \pm 0.4}$ was consistent with the hypothesis that
all LMXBs were created within globular clusters. However,
the uncertainty of the power-law exponent was large enough that a significant
field population of LMXBs could not be excluded.
The poor spatial resolution of {\it ASCA} did not allow individual LMXBs to
be resolved and their luminosities added. Instead, $L_X$ was estimated from
performing a two-component spectral fit to simultaneously model the hot gas
and LMXB components of the X-ray spectra. Such a method generally leads to
large uncertainties in the flux of the LMXB component, especially for galaxies
with much hot gas, since $\chi^2$ fitting will attempt to fit the many soft
energy channels at the expense of the fewer hard energy channels that constrain
the LMXB component the best.

The ability of {\it Chandra} to resolve a large fraction of the LMXB population
of early-type galaxies greatly reduces the uncertainties in $L_X$, which
allows better constraints to be put on the $L_X/L_{opt}$ versus $S_N$ relation.
Kim \& Fabbiano (2004) have performed such a study with 14 early-type galaxies
observed with {\it Chandra} and confirmed the result of White et al.\ (2002)
of a connection between $L_X/L_{opt}$ and $S_N$. They did not, however,
derive a slope between the two quantities.
In this paper, we analyze the {\it Chandra} data for 12 early-type galaxies
to determine the exact dependence of $L_X/L_{opt}$ on $S_N$ to put constraints
on the relative fraction of LMXBs created within globular clusters and
in the field.

Our method for determining the $L_X/L_{opt}$ versus $S_N$ relation differs from
previous studies and is presented in detail in \S~\ref{sec:method}. The
inclusion of the high-$S_N$ galaxy NGC~1399 in our sample requires special
attention in order to accurately determine its $L_X/L_{opt}$ value
and is discussed in
\S~\ref{sec:ngc1399}. Our best-fit $L_X/L_{opt}$ versus $S_N$ relation is given
in \S~\ref{sec:lxlb}, and a comparison of the predicted initial fraction
of LMXBs within globular to their present-day observed values is given in
\S~\ref{sec:fractions}. Finally, we discuss our results in
\S~\ref{sec:discussion}.

\section{The Method} \label{sec:method}

Our method for determining the dependence of $L_X/L_{opt}$ on $S_N$ differs
from previous methods in three important ways. First, we do not assume a simple
power-law fit between the two quantities. If some LMXBs are truly formed in
the field, we would expect $L_X/L_{opt}$ from these LMXBs to be a constant
from galaxy to galaxy (assuming similar ages and stellar populations of the
galaxies), while $L_X/L_{opt}$ from LMXBs within globular clusters should
scale linearly with $S_N$:
\begin{equation}
\small
(L_X/L_{opt})_{tot} = (L_X/L_{opt})_{GC} + (L_X/L_{opt})_{Field}
= A * S_N + B
\label{eq:lxlb}
\end{equation}
where $A$ and $B$ are constants. Figure~\ref{fig:models} illustrates the shape
of the $L_X/L_{opt}$ versus $S_N$ relation for a variety of scenarios. If all
LMXBs were created in the field (case~1), there would be no dependence of
$L_X/L_{opt}$ on $S_N$, so $A=0$ and $B$ is a positive constant. Conversely,
if all LMXBs are formed within globular clusters (case~2), $L_X/L_{opt}$ would
vary linearly with $S_N$, for which $A$ is a positive constant and $B=0$ (i.e.,
a galaxy with no globular clusters would not contain any LMXBs).
Finally, if LMXBs were formed both within globular clusters and the field
(case~3), $A$ and $B$ would both be positive constants. In this event,
field LMXBs would dominate in low-$S_N$ galaxies, while globular cluster LMXBs
would dominate in high-$S_N$ galaxies. Although case~1
has been definitively ruled out by the finding that 20\%--70\% of LMXBs are
presently coincident with globular clusters, neither case~2 nor case~3 has
been ruled out by previous studies. Note that the difference in the shape of
the $L_X/L_{opt}$ versus $S_N$ relation between case~2 and case~3 holds even if
a significant fraction (or all, for that matter)
of the LMXBs escaped from globular clusters into the field over the lifetime of
the galaxy. In other words, LMXBs escaping from globular clusters cannot make
a case~2 scenario look like a case~3 scenario, provided that large numbers of
globular clusters are not completely destroyed.
\centerline{\null}
\vskip3.10truein
\includegraphics{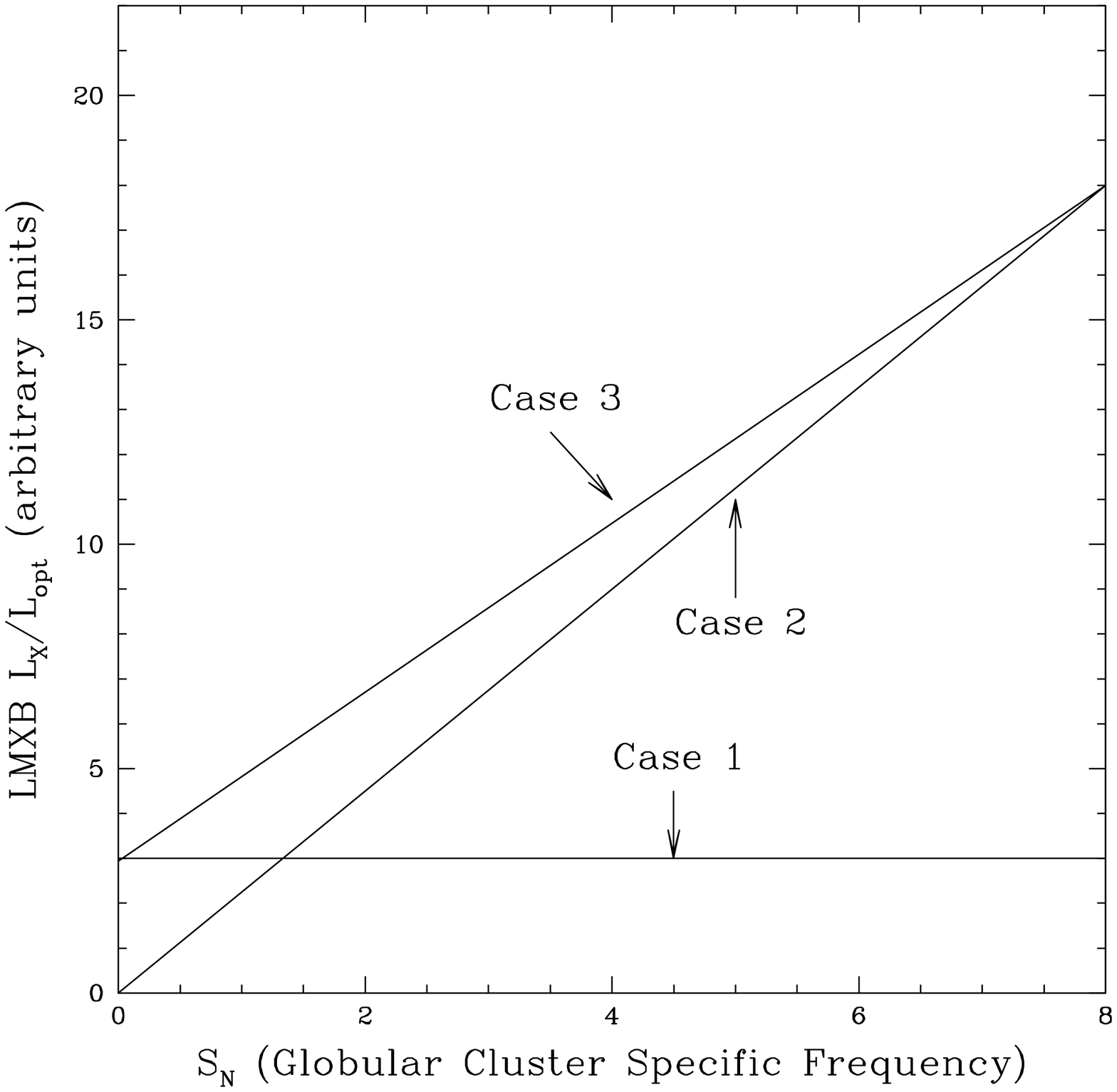}
\figcaption{ \small
The predicted relations between $L_X/L_{opt}$ and the globular cluster
specific frequency $S_N$ if all LMXBs are formed in the field (case~1),
if all LMXBs are formed within globular clusters (case~2), and if LMXBs
formed in both the field and within globular clusters (case~3).
\label{fig:models}}
\normalsize
\vspace{0.1truein}

Second, it is vitally important to calculate $L_X/L_{opt}$ only
within the same angular area for which $S_N$ was determined. It is
well established that $S_N$ is a strong function of galactic radius,
as the integrated
$S_N$ determined out to a radius of $10^{\prime}$ or more from ground-based
observations can be as much as a factor of 5 higher than when determined
within the {\it Hubble Space Telescope (HST)} Wide Field Planetary Camera 2
(WFPC2) field of view (Kundu \& Whitmore 2001a,b;
Kissler-Patig 1997).  Comparing $L_X/L_{opt}$ which was determined over the
entire {\it Chandra} ACIS-S $8^{\prime} \times 8^{\prime}$ field  of view to,
$S_N$, determined either in the $150^{\prime\prime}$ chevron-shaped field of
view of the {\it HST} WFPC2 or from large angular ($>10^{\prime}$ radius) area
ground-based observations, will undoubtedly lead to erroneous
results given the strong dependence of $S_N$ on galactic radius.

Third, in determining the total X-ray luminosity from the sum of the sources,
we exclude sources more luminous than $5\times10^{38}$ ergs s$^{-1}$. The few
most luminous sources in a typical early-type galaxy generally contribute
a significant fraction of the total X-ray luminosity from LMXBs. Thus, the
determination of $L_{X,total}$ can be greatly affected by random Poisson
fluctuations in the number of the few most luminous LMXBs,
especially for smaller galaxies
that are only expected to have two to three sources with $L_X>5\times10^{38}$
ergs s$^{-1}$. In fact, for the sample of galaxies we study here, we find that
the fraction of the total flux emanating from $>5\times10^{38}$ ergs s$^{-1}$
sources ranges anywhere from 0\% to 35\%. However, by excluding sources more
luminous than $5\times10^{38}$ ergs s$^{-1}$ and only including galaxies large
enough such that the integrated luminosity from sources below $5\times10^{38}$
ergs s$^{-1}$ is at least $5\times10^{39}$ ergs s$^{-1}$, no one source
will contribute more than 10\% to the total flux, thereby minimizing the
effect of an overabundance or deficit of high-luminosity sources
on $L_{X,total}$.

\section{Observations and Data Reduction} \label{sec:observations}

We constructed a sample of every normal early-type galaxy that satisfied the
following criteria: (1) it was observed by the {\it Chandra} ACIS-S for at least
10 ks, (2) the literature $S_N$ value was constrained to at least 50\%
accuracy and the value of the $V$-band luminosity each author used to
determine $S_N$ over the reported angular area is stated, (3) the total
LMXB luminosity was at least $5 \times 10^{39}$ ergs s$^{-1}$ (see above),
and (4) it did not contain so much hot gas that determination of the unresolved
LMXB emission would be problematic, unless literature $S_N$ values were
available in a well-defined angular region outside the core of the galaxy (such
as NGC~1399 below). This yielded a total of 12 galaxies, listed in
Table~\ref{tab:galaxies}.
 
The data for each galaxy were processed in a uniform manner following the
{\it Chandra} data reduction threads.
The data were calibrated with the most recent gain maps
at the time of reduction.
Pile-up was not an issue even for the brightest sources and no
correction has been applied.
Energy channels above 6 keV were ignored, since these channels introduce
significantly more noise than source signal.
Sources were detected using the ``Mexican-Hat" wavelet detection routine
{\it wavdetect} in CIAO in an 0.3--6.0 keV band image. We then culled the
source list to include only those sources that were detected at the $>3\sigma$
level. All images created were corrected for exposure and vignetting.
For each source a local background was determined from a circular annular
region with an inner radius that was set to 1.5 times the semimajor axis
of the source extraction region and an outer
radius chosen such that the area of the background annulus was 5
times the area of the sources extraction region. Care was taken to exclude
neighboring sources from the background annuli of each source in crowded
regions. For the diffuse emission, background was taken from the same S3 chip
well away from the detectable galactic emission.

For our study, we only included sources that fell within the angular area
covered by the optical telescope used to determine $S_N$. For all galaxies
except NGC~1399, this was the familiar chevron-shaped {\it HST} WFPC2 field of
view. For NGC~1399, we considered sources within two separate
annular rings that were $30^{\prime\prime}-90^{\prime\prime}$ and
$90^{\prime\prime}-150^{\prime\prime}$
in extent, for comparison to $S_N$ determined in these annular bins by
Ostrov, Forte, \& Geisler (1998) from ground-based observations.
For each source, the 0.3--6.0 keV count rate was converted into a flux using
a $\Gamma=1.56$ power-law model, a value typical of the integrated flux from
LMXBs in early-type galaxies (Irwin et al.\ 2003).
Unabsorbed fluxes were then determined
assuming Galactic hydrogen column densities from Dickey \& Lockman (1990), and
converted to 0.3--10 keV luminosities assuming surface brightness fluctuation
distances of Tonry et al.\ (2001). The luminosities of each of the sources
except the central source (which might be a low-luminosity active galactic
nucleus [AGN] rather than an
LMXB) were summed to yield $L_{X, resolved}$ for each galaxy.
To this we added an estimate of
the total luminosity of unresolved LMXBs that were either below the detection
threshold of the observation or for which crowding at the center of the galaxy
precluded their individual detection. To estimate $L_{X, unresolved}$, we
tried performing spectral fitting of the diffuse emission to simultaneously
model the hot gas (APEC) and unresolved LMXB (power-law) components
within XSPEC, but doing so consistently led to a best-fit model that
underestimated the 2--6 keV flux by 10\%-20\%. Evidently, complexities in
modeling the hot gas (the effects of temperature gradients and non-solar
abundance ratios) have led the fitting routine to attempt to minimize $\chi^2$
in the numerous, gas-dominated soft energy channels at the expense of the fewer
LMXB-dominated hard energy channels. We therefore use an alternative method to
determine $L_{X, unresolved}$ in which
we assume that the diffuse emission above a certain energy is not contaminated
by emission from the hot gas in these galaxies, leaving unresolved LMXBs as the
sole source of the diffuse emission. Gas at a temperature of 0.7 keV typically
contributes a negligible number of counts above 2 keV given the {\it Chandra}
response, while gas at 0.3 keV contributes negligibly above 1 keV. Starting
with the 3--6 keV background-subtracted diffuse flux,
we estimated the 0.3--10 keV luminosity of unresolved sources assuming the same
$\Gamma=1.56$ power-law model as was assumed for the resolved sources. We then
stepped down the low-energy bound of the diffuse flux determination in 0.5 keV
increments until the extrapolation to the 0.3--10 keV band failed to yield a
value consistent with the extrapolation from using higher low energy cut offs
(indicating that hot gas was contributing to the diffuse flux) and calculated
the unresolved LMXB luminosity using a lower energy cut off that was one
0.5 keV increment higher. This estimate for $L_{X, unresolved}$ was added to
$L_{X, resolved}$ to yield $L_{X, total}$ for all sources in each galaxy.
For most of the galaxies in the sample the resolved source flux represented
between 25\% to 50\% of the total LMXB flux. The 1~$\sigma$ uncertainty in the
luminosity was calculated accordingly from the total number of counts for each
galaxy and ranged from 7\% to 20\%.

A fraction (typically 5\%) of the observed X-ray flux emanates from unrelated
foreground/background sources along the line of sight to the galaxy. Using the
log $N$--log $S$ relation of Mushotzky et al.\ (2000), we have determined the
expected number and integrated flux of such sources over the angular area
studied, and subtracted this flux from the observed flux.

The foreground/background source-corrected $L_{X, total}$ was normalized
by the optical light of the galaxy that fell within our angular region of
interest. We used the same $V$-band luminosities from the literature that
provided the values of $S_N$. Historically, $L_X/L_{opt}$ has been given in
terms of the $B$-band luminosity ($L_X/L_B$), so for comparison to previous
studies, we convert the $V$-band luminosities to $B$-band using published $B-V$
values. Since all the galaxies in our sample have nearly identical
$B-V$ values, the choice to use $B$-band luminosities rather than
$V$-band luminosities has a minimal impact on the results. 

\section{The Case of NGC~1399} \label{sec:ngc1399}

Since most of the galaxies in our sample have globular cluster specific
frequencies at or below 2.0, the addition of a galaxy with a very high $S_N$
value such as NGC~1399 would considerably improve the constraint on the
$L_X/L_B$ versus $S_N$ relation. Unfortunately, the hot gas within NGC~1399
is about 1 keV. At this temperature and given the large amount of gas that
NGC~1399 contains, the gas is still contributing appreciably to the flux at
3 keV. In addition, the gaseous flux from NGC~1399 fills the entire S3 chip,
making background subtraction problematic.
As a result, our method of estimating the unresolved LMXB emission
from the amount of diffuse emission in the hard energy band will not work.

An alternative way of estimating the integrated LMXB luminosity from NGC~1399
is to use the fact that the X-ray luminosity functions of the sources within
early-type galaxies are fairly regular, at least for sources less luminous
than $5 \times 10^{38}$ ergs s$^{-1}$. From the other galaxies in our sample
for which the detection limit was below $10^{38}$ ergs s$^{-1}$,
we calculated the fraction of the total integrated flux emanating from sources
with luminosities in the range $(1-5) \times 10^{38}$ ergs s$^{-1}$, i.e.,
$f = \frac{\displaystyle\sum L_{X,i}~(1-5 \times 10^{38} \rm ergs~s^{-1})}
{\displaystyle\sum L_{X,i}~(<5 \times 10^{38} \rm ergs~s^{-1})}$, and
found that $f$ was reasonably similar from galaxy to galaxy, ranging from
34\% to 48\%, with a mean of 41\% and a standard
deviation of 5\%. We then determined the luminosity from resolved sources
in the range $(1-5) \times 10^{38}$ ergs s$^{-1}$ for NGC~1399 and divided this
luminosity by $0.41\pm0.05$ to yield the total luminosity from LMXBs in
NGC~1399. This was done separately for two annular rings
$30^{\prime\prime}-90^{\prime\prime}$ and
$90^{\prime\prime}-150^{\prime\prime}$. The detection limit for this observation
of NGC~1399 was $7\times 10^{37}$ ergs s$^{-1}$, so we can be reasonably
assured of detecting all sources above $10^{38}$ ergs s$^{-1}$ outside of
$30^{\prime\prime}$. For the inner $30^{\prime\prime}$
the high X-ray surface brightness of the hot gas precluded the detection of
all but the most luminous sources, rendering this technique ineffective for
this region.

\section{The $L_X/L_B$ Versus $S_N$ Relation} \label{sec:lxlb}

The $L_X/L_B$ versus $S_N$ relation for the galaxies in our sample is shown in
Figure~{\ref{fig:lxlb}. We fit a linear model of the form
$L_X/L_B = A * S_N + B$ to the data using a linear regression algorithm
that accounts for errors in both quantities (see Fasano \& Vio 1988). Here
$L_X/L_B$ is in units of $10^{29}$ ergs s$^{-1}$ L$_{B_{\odot}}^{-1}$.
Without including the two data points from NGC~1399, the best-fit
coefficients are $A=2.72\pm0.47$ and $B=2.87\pm0.50$ (1 $\sigma$ uncertainties).
Note that the two data
points for NGC~1399 lie near the best-fit relation, indicating that
our alternative technique for determining $L_X/L_B$ for NGC~1399 is reasonable.
If we include NGC~1399, the best-fit constants are
$A=2.41\pm0.33$ and $B=3.12\pm0.39$.

The best-fit relation between $L_X/L_B$ and $S_N$ most closely resembles the
prediction of case~3 shown in Figure~\ref{fig:models}. Here $B$ differs from 0
at
the $8.0 \sigma$ level and strongly argues that a significant fraction of LMXBs
were truly formed in the field, rather than having been formed within globular
clusters and later escaped. If the latter had occurred, we would still
expect a linear relation through the origin (case~2).

\centerline{\null}
\vskip3.10truein
\includegraphics{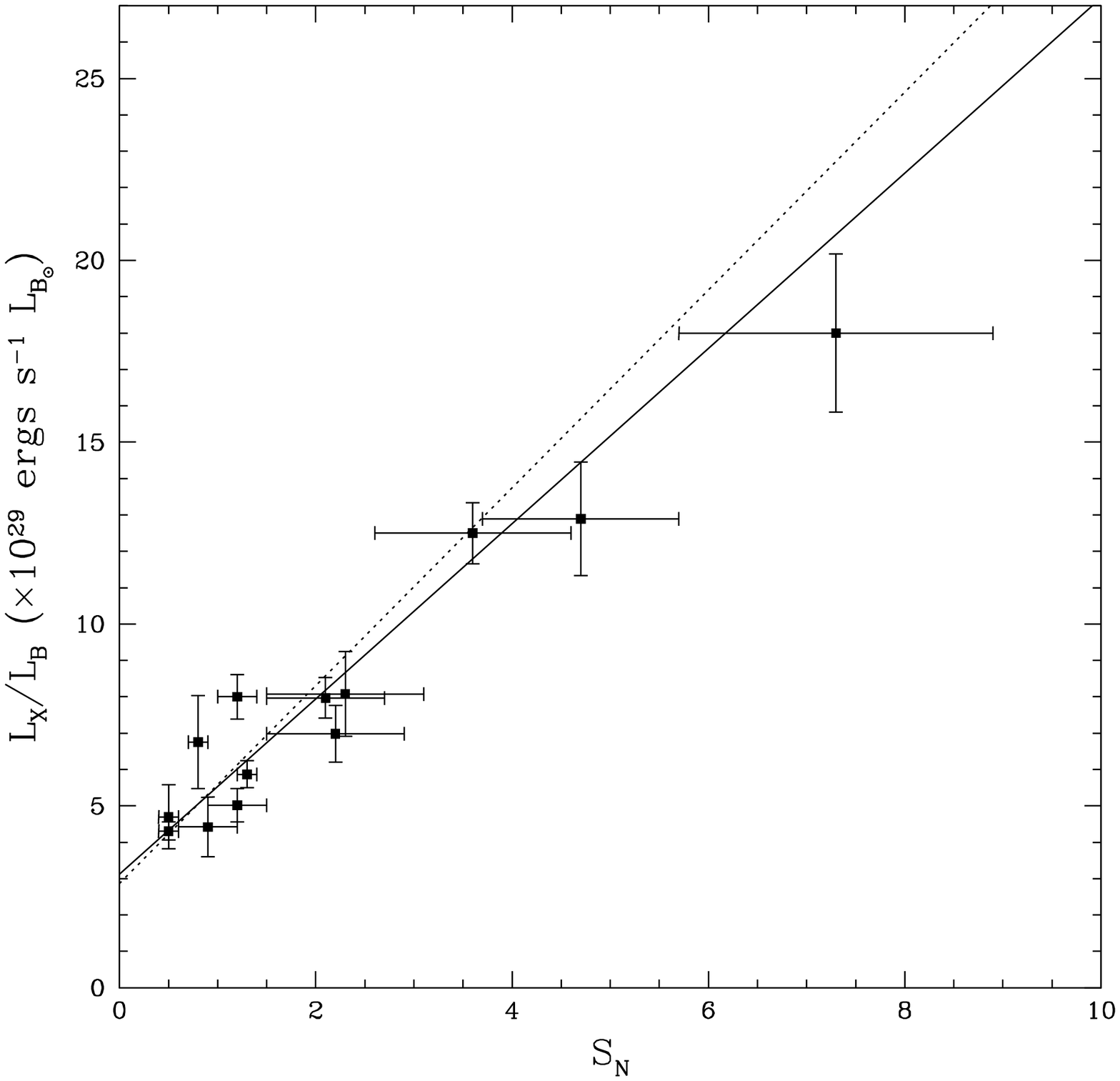}
\figcaption{ \small
The {$L_X/L_B$} relation for the 12 galaxies in our sample
(with two data points from NGC~1399) with 1$\sigma$ uncertainties for both
quantities. The dotted line represents the best-fit relation if
NGC~1399 is omitted, while the solid line represents the best-fit relation
if NGC~1399 is included in the fit. In both cases, the $y$-intercept of the
relation is clearly different from 0, indicating that some of the LMXBs are
formed in the field.
\label{fig:lxlb}}
\normalsize
\vspace{0.1truein}

One potential cause of bias in a study of this nature stems from the fact that
LMXBs are observed to occur in red globular clusters 3 times more often
than in blue globular clusters, both within the Milky Way (Grindlay 1993)
and in nearby elliptical galaxies (Kundu et al.\ 2002; Sarazin
et al.\ 2003). A galaxy with a higher fraction of red globular clusters would
create more LMXBs than a galaxy with a lower fraction of red globular clusters,
even if both galaxies had the same total number of globular clusters. This
would create scatter in the $L_X/L_B$ versus $S_N$ relation that might bias
the results drawn from it. To test whether this has affected our results,
we determined the fraction of red clusters in each galaxy in our sample, using
the $V-I$ histograms of the globular clusters for each galaxy published in the
literature
(Kundu \& Whitmore 1998; 2001a,2001b). We considered a globular cluster to be
blue if its $(V-I)_0<1.1$, and red if its $(V-I)_0>1.1$ (following Kundu
et al.\ 2002). The observed colors
were corrected for Galactic extinction. For the galaxies in our sample
the red fraction of globular clusters varied from 24\% to 48\%. Previous
studies have found that red clusters have a $\sim$6\% chance of harboring
a $>10^{37}$ ergs s$^{-1}$ LMXB, while blue clusters have a $\sim$2\% chance.
Given these fractions, we found that a galaxy with a red globular cluster
fraction of 48\% will produce about 30\% more LMXBs than a galaxy with a
red globular cluster fraction of only 24\%, neglecting the presence of field
LMXBs. If field LMXBs are included and are assumed to comprise roughly half
of the total LMXBs in a galaxy, the difference in the total number of LMXBs
(and hence $L_X/L_B$) between the two galaxies diminishes to 15\%, or
$\pm8\%$ from the mean. This is on the order of the statistical uncertainties
of each galaxy in our sample. Furthermore,
the red fraction of globular clusters is uncorrelated with $S_N$ for the
galaxies in our sample, eliminating the possibility that this effect
introduces any systematic bias to the $L_X/L_B$ versus $S_N$ relation.
Thus, while a varying fraction of red clusters
in a galaxy can introduce some scatter in the $L_X/L_B$ versus $S_N$ relation,
it is not enough to change the overall results of this study.

Another possible source of bias is if a non-negligible fraction of the unresolved
hard X-ray flux emanated from a large population of weak X-ray emitters
(i.e., RS CVn systems, cataclysmic variable, M stars, or some other unknown
X-ray emitter) rather than LMXBs. If the summed X-ray luminosity of such sources
scales linearly with the mass of the galaxy, then this would account for at
least part of the derived $(L_X/L_B)_{field}$ value that we had assumed to be
exclusively from LMXBs, and the removal of such a component from the
$L_X/L_B$ versus $S_N$ relation would diminish the statistical need for a
field-born population of LMXBs (i.e., $B$ would be lowered and less
constrained). To investigate this, we have analyzed an archival 38 ks
{\it Chandra} observation of the bulge of M31, which we assume can serve
as an adequate template for the X-ray population of old stellar populations like
those within the galaxies of our sample. We found that 86\% of the 2--6 keV
flux was resolved into point sources with luminosities exceeding $1.5 \times
10^{36}$ ergs s$^{-1}$. Thus, if there is a hidden population of weak X-ray
emitters within old stellar populations, they contribute less than 14\%
to the hard X-ray flux, or equivalently, less than 14\% of our best-fit $B$
value.

It should be noted that a non-zero $y$-intercept implies that all or most
LMXBs in the field formed in the field only if we assume that a significant
number of globular clusters were not destroyed after the ejection of the
LMXB (or if the LMXB survived the process that destroyed the globular
cluster). If many globular clusters were destroyed, this would have the effect
of reducing $S_N$ while leaving $L_X/L_B$ unchanged, shifting the true
$L_X/L_B$ versus $S_N$ relation to the left. Thus, it might be possible that
if all LMXBs were formed within globular clusters (case~2), the present-day
$L_X/L_B$ versus $S_N$ relation could resemble case~3 if enough globular
clusters were destroyed. However, this cannot be accomplished by simply
destroying the same fraction of globular clusters in each galaxy
(dividing the original $S_N$ of each galaxy by a constant), as this would still
lead to a best-fit relation through the origin which is not seen. What would
be required is that some galaxies have lost a much higher fraction of their
globular clusters than other galaxies. From Figure~{\ref{fig:lxlb}, even if
we assume that the galaxies with $S_N > 2$ have not lost any globular clusters,
it would require that the two lowest $S_N$ galaxies (NGC~1553 and NGC~3379)
have lost about two-thirds of their globular clusters (i.e., they had
an original $S_N$ of $\sim$1.5) if the original $L_X/L_B$ versus $S_N$ relation
were to go through the origin. While the globular cluster destruction rates
of Vesperini (2000) predict such a high destruction rate for galaxies less
massive than $10^{10}$ M$_{\odot}$, NGC~1553 and NGC~3379 have masses
approaching 
$10^{12}$ M$_{\odot}$, given their absolute visual magnitudes and a reasonable
$M/L$ ratio. Vesperini (2000) predicts that galaxies of this mass should only
lose about 10\% of their original globular cluster population.
Furthermore, NGC~1553 and NGC~3379 have absolute visual magnitudes comparable
to as the $S_N > 2$ galaxies (Table~\ref{tab:galaxies}),
so it is unclear why these galaxies should lose such a high fraction of their
globular clusters while the other galaxies do not.
Finally, the destruction of such a large number of globular clusters in certain
galaxies would lead to much more pollution of LMXBs into the field than in
galaxies that lost few globular clusters. This would lead to a large dispersion
in the measured $L_X/L_B$ values of the field LMXBs. In Figure~\ref{fig:field}}
we show the $L_X/L_B$ versus $S_N$ relation where $L_X/L_B$ has been broken into
its {\it observed} field and globular cluster
components for the galaxies in our sample for which observed fractions
of LMXBs within globular clusters appear in the literature (and listed in the
fourth column of Table~\ref{tab:compare}}). Whereas $(L_X/L_B)_{globular}$ is
a strong function of $S_N$, $(L_X/L_B)_{field}$ is not any higher in the
$S_N=0.5$ galaxy than it is for $S_N>2$ galaxies, indicating that it has not
undergone significantly more pollution than galaxies with higher $S_N$.
\centerline{\null}
\vskip3.10truein
\includegraphics{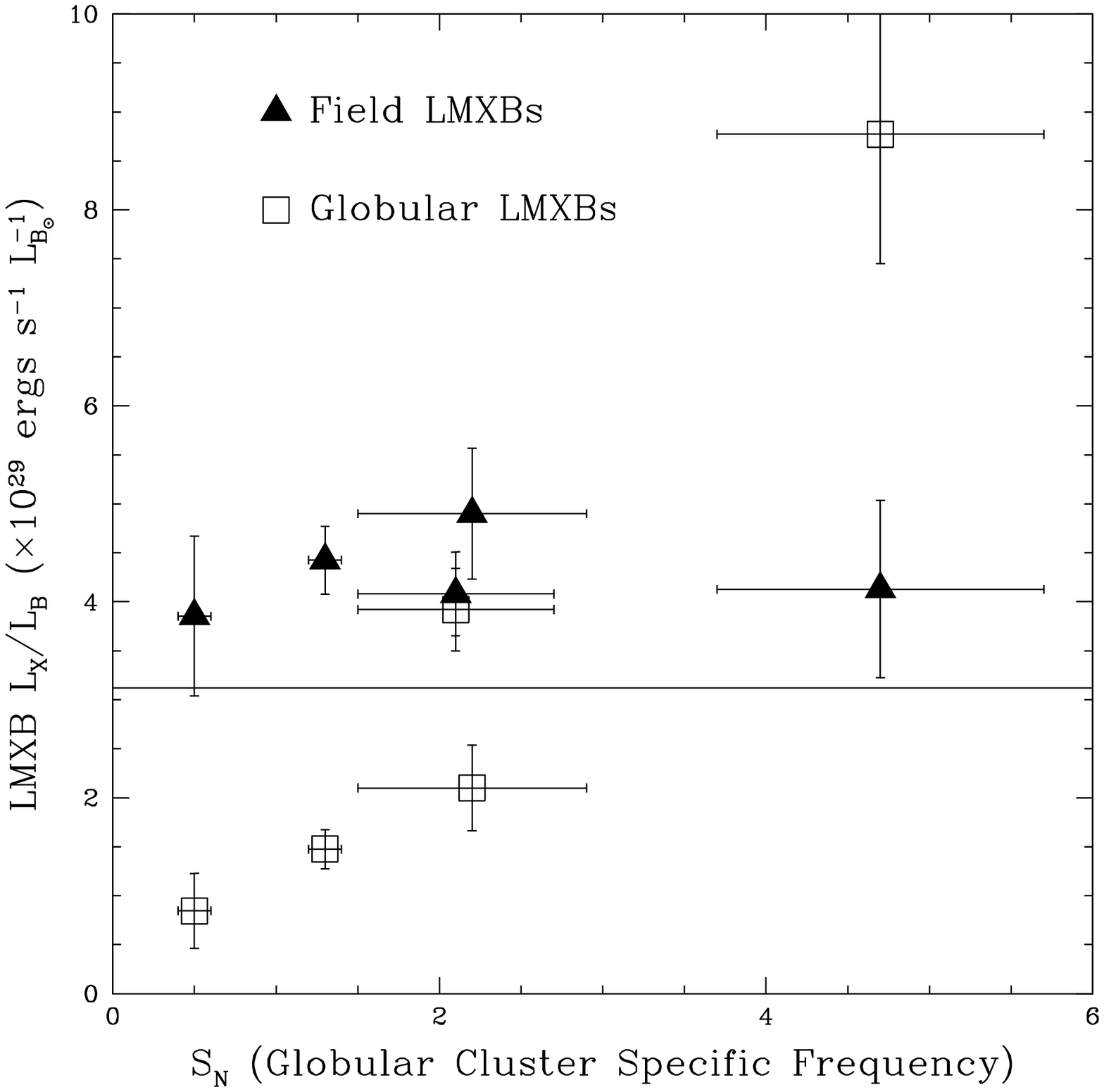}
\figcaption{ \small
The {$L_X/L_B$} relation for which {$L_X/L_B$} has been
broken into its {\it observed} globular cluster and field components for
NGC~1553, NGC~3115, NGC~4365, NGC~1332, and NGC~1399. The solid
horizontal line represents the predicted initial field {$L_X/L_B$} value.
The fact that the data points lie above this line (especially the S0
galaxies NGC~3115 and NGC~1332) indicates that the field has been polluted
to some extent by LMXBs that have escaped from globular clusters.
\label{fig:field}}
\normalsize
\vspace{0.1truein}

\section{The Initial and Present Day Fractions of LMXBs Within Globular
Clusters} \label{sec:fractions}

The non-zero $y$-intercept of the $L_X/L_B$ versus $S_N$ relation demonstrates
that LMXBs are formed within the field, but it does not eliminate the
possibility that at least some of the present-day field population was
ejected from globular clusters. If some of the present-day field LMXBs actually
were created in and later escaped from globular clusters, we would expect
the present-day fraction of LMXBs within globular clusters to be smaller than
the initial fraction of LMXBs within globular clusters. The latter quantity
can be predicted from our $L_X/L_B$ versus $S_N$ relation and can be given in
terms of the best-fit constants $A$ and $B$ along with $S_N$ for each galaxy:
\begin{equation}
{\rm Initial~\%~in~GCs}~= \frac{(L_X/L_B)_{GC}}{(L_X/L_B)_{Total}} \\
= \frac{A * S_N}{A * S_N + B}.
\label{eq:fraction}
\end{equation}
It is important to keep in mind that nowhere in the determination of the
$L_X/L_B$ versus $S_N$ relation (and therefore the determination of the initial
fraction of LMXBs within globular clusters) did we use the knowledge of the
{\it observed} (present-day) fraction of LMXBs within globular clusters.
Thus, the
comparison of the initial fraction of LMXBs formed within globular cluster
to the present-day observed fraction is not tautological in nature.
In Table~\ref{tab:compare} we compare the initial and measured fractions of
LMXBs formed within globular clusters for those galaxies
for which such a measurement exists in the literature. We have also included
NGC~4472 and NGC~4649, 2 galaxies we excluded from our original sample
because they contain large amounts of $\sim$1 keV gas that prevents an
accurate determination of the unresolved LMXB flux. There is good agreement
between the initial and measured fractions of LMXBs within globular clusters
for the four elliptical galaxies (NGC~4472, NGC~4649, NGC~4365, and NGC~1399),
where the measured fraction is only slightly lower than the initial fraction,
indicating that relatively few LMXBs have escaped from
globular clusters in these
galaxies. However, for the three S0 galaxies (NGC~1553, NGC~3115, and NGC~1332)
the measured fraction is about a factor of
two less than the initial fraction, indicating that a significant number of
globular cluster LMXBs have escaped into the field in these galaxies.
This is illustrated further in Figure~\ref{fig:field}, where it can be seen
that the field $L_X/L_B$ values of the S0 galaxies NGC~3115 and NGC~1332
(NGC~1553 has large statistical uncertainties) are
several sigma above the best-fit predicted initial $L_X/L_B$ value of
$3.12 \times 10^{29}$ ergs s$^{-1}$ L$_{B_{\odot}}^{-1}$
(our best-fit constant $B$ from \S~\ref{sec:lxlb}), with the difference being
from LMXBs ejected from globular clusters into the field. Such an effect is
seen to a lesser extent in the elliptical galaxies NGC~4365 and NGC~1399,
where fewer LMXBs were ejected from LMXBs.

\section{Discussion} \label{sec:discussion}

The shape of the $L_X/L_B$ versus $S_N$ relation for the 12 early-type galaxies
in our sample strongly argues that the number of LMXBs in a galaxy does not
scale directly with the number of globular clusters the galaxy contains. We
have demonstrated that the non zero $y$-intercept of the
$L_X/L_B$ versus $S_N$ relation is not the result of large-scale globular
cluster destruction or a variation in the red--to--blue
globular cluster ratio but instead can only be explained in terms of a
population of LMXBs that formed within the field.

These field LMXBs are most likely primordial binary systems formed during the
the last major star formation episode of the galaxies 5--12 billion years ago.
Obviously, these sources cannot be persistent X-ray emitters, since the mass
accretion rates needed to power the observed X-ray luminosities would consume
all the mass of a $<1$ M$_{\odot}$ donor star in $<10^8$ yr. The onset of
the X-ray active phase of the LMXB may be delayed by several billion years in
wider binary systems if the donor star is required to enter its red giant phase
in order to overflow its Roche lobe. In addition, wide binaries are expected
to be transient in nature, with LMXBs with periods of $>1$ day expected to be
in their ``off" stage $>75\%$ of the time (Piro \& Bildsten 2002), further
lengthening the lifetime of an LMXB system. On the other hand, LMXBs
within globular clusters are most likely short-period systems owing to the
fact that frequent stellar encounters will serve to tighten the orbit of an
existing binary, leading to persistent X-ray emission in these systems (again
from Piro \& Bildsten 2002). This expected transient nature of long-period
primordial LMXBs and persistent nature of short-period dynamically-created
LMXBs could be used as a means of confirming that present-day field LMXBs
were created in the field if multi-epoch X-ray observations find
that field LMXBs are transient whereas globular cluster LMXBs are persistent
emitters. The largely transient nature of field LMXBs would also imply
that the efficiency of creating LMXBs in the field is much higher than
previously thought, since at any given time we are only observing the small
fraction of the total field LMXB population that happen to be in their ``on"
($L_X > 10^{37}$ ergs s$^{-1}$) state.

Given that $S_N$ generally increases with galactic radius (Ashman \& Zepf
1998), Equation~\ref{eq:lxlb} predicts that $L_X/L_B$ should increase
with increasing galactic radius. This is in fact seen in NGC~1399, for
which the measured $L_X/L_B$ increases by about 50\% as $S_N$ increases
from 4.7 in the $30^{\prime\prime}-90^{\prime\prime}$ annular bin to
7.3 in the $90^{\prime\prime}-150^{\prime\prime}$ annular bin. The fraction
of LMXBs found within globular clusters should also increase as a function
of radius as the number of globular cluster LMXBs increases (per unit light),
whereas the number of field LMXBs remains constant (also per unit light).
Deeper {\it Chandra} observations will test this prediction, although
considerable care will be needed to account for the fact that contamination
by background AGN is more prevalent at larger radii, and failure to account for
this will lead to an underestimate of the true fraction
of LMXBs within globular clusters. Also,
the effects of possible radial gradients in the red/blue ratio of globular
clusters will need to be properly accounted for in determining the
fraction of LMXBs within globular clusters as a function of radius.

Another key result of this study is that the fraction of LMXBs currently
within globular clusters of S0 galaxies is only half of the initial fraction
of LMXBs within globular clusters predicted from the $L_X/L_B$ versus $S_N$
relation. This effect is not seen in the elliptical galaxies in our sample,
for which the predicted initial fractions of LMXBs within globular clusters
is consistent with (albeit slightly lower than) the measured fractions.
It is unlikely that the higher than expected fraction of field LMXBs in S0s
is the result of LMXBs forming more efficiently in the fields of S0s than in
ellipticals, since the required number of ``extra" field LMXBs would lead to
$(L_X/L_B)_{Total}$ values for S0s that are a factor of 2 greater than in
elliptical galaxies with the same $S_N$, and $(L_X/L_B)_{field}$ values that
are 3 times higher. Previous studies have indicated that there are no
discernible differences in the properties of globular clusters within S0 and
elliptical galaxies that might account for the increased escape rate of LMXBs
from globular clusters within S0 galaxies. Kundu \& Whitmore (2001a,2001b) found
that mean colors, mean half-light radii, and luminosity functions of the
globular
clusters found in samples of S0 and elliptical galaxies were statistically
identical. This would argue that the LMXB ejection mechanism does not
involve processes internal to the clusters such as supernova kicks.
Another reason why supernova kicks are not favored is that such a kick
would remove the neutron star or black hole from the globular cluster only
a few million years after the formation of the cluster. Spending such a short
period of time within the cluster would eliminate the advantage that globular
clusters have in creating LMXBs (Maccarone et al.\ 2003),
meaning that these LMXBs would effectively be primordial binaries too.
Instead, it is more likely that LMXBs within globular clusters of S0 galaxies
have been removed by the tidal disruption (but not destruction)
of globular clusters. This effect is expected to be
more pronounced within S0 galaxies than in elliptical galaxies, owing to the
presence of disks in S0s that are lacking in ellipticals. Gravitational shocks
caused by the passage of a globular cluster through a disk are much
stronger than passage through a bulge distribution (Fall \& Zhang 2001),
leading to a greater number of LMXBs ejected from globular clusters in S0
galaxies than in ellipticals.
Clearly, more observations are needed to confirm this hypothesis. If the field
LMXBs in S0s are composed of a mix of primordial binaries and LMXBs ejected
from globular clusters, then the former should be transient X-ray emitters
while the latter should be persistent emitters. On the other hand, field
LMXBs within elliptical galaxies should nearly all be transient.
Once again, multi-epoch
X-ray observations of the field LMXBs could confirm this picture if the
predicted ratio of persistent--to--transient field LMXBs is measured.

\acknowledgements
We thank Joel Bregman, Renato Dupke, and Chris Mullis for
illuminating discussions. We also thank an anonymous referee for many
useful suggestions and Fionn Murtagh for
providing us with his linear regression algorithm adopted from Fasano \&
Vio.
Support for this work was provided by the National Aeronautics and Space
Administration through Chandra Award Number GO4-5097X issued by the {\it Chandra
X-ray Observatory} Center, which is operated by the Smithsonian Astrophysical
Observatory for and on behalf of the National Aeronautics Space Administration
under contract NAS8-03060.
This research has made use of the NASA/IPAC Extragalactic Database (NED),
which is operated by the Jet Propulsion Laboratory, California Institute
of Technology, under contract with NASA.

\small
\begin{deluxetable}{lccccccccc}
\tablecaption{{\it Chandra} Sample of Galaxies}
\tablewidth{0pt}
\tablehead{
\colhead{Galaxy} & \colhead{Type} & \colhead{Distance} & $M_V$ &
\colhead{$L_X/L_B$\tablenotemark{a}} & \colhead{$S_N$} & 
\colhead{$L_{X,limit}$} & \colhead{$C_{res}$\tablenotemark{f}} &
\colhead{$C_{unres}$\tablenotemark{g}} & \colhead{$C_{gas}$\tablenotemark{h}} \\
 &  & \colhead{(Mpc)} & &   &  &
\colhead{(ergs s$^{-1}$)}  & & &
}
\startdata
NGC~1553 & S0 & 18.5 & -22.0 & $4.7 \pm 0.9$
& $0.5 \pm 0.1$\tablenotemark{b}~~&  $1.8\times10^{38}$ & 122 & 567 & 1153 \\
NGC~3379 & E1 & 10.6 & -20.9 & $4.3 \pm 0.3$
& $0.5 \pm 0.1$\tablenotemark{c}~~&  $4.1\times10^{37}$ & 598 & 383 & 208 \\
NGC~4406 & S0/E3 & 17.1  & -22.1 & $6.8 \pm 1.3$
& $0.8 \pm 0.1$\tablenotemark{c}~~&  $2.6\times10^{38}$ & 93 & 554 & 4320  \\
NGC~4494 & E1 & 17.1 & -21.5 & $4.4 \pm 0.8$
& $0.9 \pm 0.3$\tablenotemark{c}~~&  $1.8\times10^{38}$ & 165 & 108 & 63  \\
NGC~4621 & E5 & 18.3 & -21.7 & $5.0 \pm 0.5$
& $1.2 \pm 0.3$\tablenotemark{c}~~&  $1.5\times10^{38}$ & 307  & 231 & 61 \\
NGC~4552 & E & 15.3 & -21.3 &  $8.0 \pm 0.6$
& $1.2 \pm 0.2$\tablenotemark{c}~~& $4.9\times10^{37}$ & 1057 & 1408 & 10,602 \\
NGC~3115 & S0 & 9.7 & -21.1 &  $5.9 \pm 0.4$
& $1.3 \pm 0.1$\tablenotemark{d}~~&  $3.0\times10^{37}$ & 931 & 637 & 292 \\
NGC~4365 & E3 & 20.4 & -22.0 & $8.0 \pm 0.6$
& $2.1 \pm 0.6$\tablenotemark{c}~~&  $1.0\times10^{38}$ & 603  & 709 & 801 \\
NGC~1332 & S0 & 22.9 & -21.5 & $7.0 \pm 0.8$
& $2.2 \pm 0.7$\tablenotemark{b}~~&  $1.4\times10^{38}$ & 354 & 893 & 3325  \\
IC~1459 & E3 & 29.2 & -22.5 & $8.1 \pm 1.2$
& $2.3 \pm 0.8$\tablenotemark{c}~~&  $1.8\times10^{38}$ & 349 & 1073 & 3047  \\
NGC~4278 & E1/2 & 16.1 & -21.0 & $12.5 \pm 0.8$
& $3.6 \pm 1.0$\tablenotemark{b}~~&  $7.6\times10^{37}$& 488 & 908 & 320 \\
NGC~1399 & E1 & 20.0 & -22.0 & $12.9 \pm 1.8$
& $4.7 \pm 1.0$\tablenotemark{e}~~&  $7.2\times10^{38}$ & 1235 & 1752 & 20,980\\
NGC~1399 &\ldots &\ldots & \ldots &$18.0 \pm 2.6$
& $7.3 \pm 1.6$\tablenotemark{e}~~&  $7.2\times10^{38}$ & 1141 & 1517 & 16,875\\
\enddata

\tablenotetext{a}{in units of $10^{29}$ ergs s$^{-1}$ L$_{B_{\odot}}^{-1}$,
where $L_X$ is in the 0.3--10 keV band and only includes sources
with individual X-ray luminosities below $5 \times 10^{38}$ ergs s$^{-1}$.
Both $L_X$ and $L_B$ are determined only within the stated field of view.}
\tablenotetext{b}{Globular cluster specific frequency from Kundu \& Whitmore
(2001a).}
\tablenotetext{c}{Globular cluster specific frequency from Kundu \& Whitmore
(2001b).}
\tablenotetext{d}{Globular cluster specific frequency from Kundu \& Whitmore
(1998).}
\tablenotetext{e}{Globular cluster specific frequency from Ostrov, Forte, \&
Geisler (1998) for $30^{\prime\prime}-90^{\prime\prime}$ and
$90^{\prime\prime}-150^{\prime\prime}$ annular regions.}
\tablenotetext{f}{Background-subtracted counts from resolved LMXBs with
$L_X < 5 \times 10^{38}$ ergs s$^{-1}$ in area of the interest.}
\tablenotetext{g}{Background-subtracted counts from unresolved LMXBs in area
of the interest.}
\tablenotetext{h}{Background-subtracted counts from diffuse gas in area of
the interest.}

\label{tab:galaxies}
\end{deluxetable}

\begin{deluxetable}{lccc}
\tablecaption{Fraction of LMXBs Within Globular Clusters}
\tablewidth{0pt}
\tablehead{
\colhead{Galaxy} & \colhead{$S_N$} & \colhead{Initial Percentage
(from Equation~\ref{eq:fraction})} & \colhead{Measured Percentage}
}
\startdata
NGC~1553 & $0.5\pm0.1$ & $28^{+4}_{-4}\%$ & ~~~$18^{+9}_{-12}\%$\tablenotemark{a} \\
NGC~4472 & $1.1\pm0.1$ & $46^{+2}_{-2}\%$ & ~~~$42^{+7}_{-6}\%$\tablenotemark{b} \\
NGC~3115 & $1.3\pm0.1$ & $50^{+2}_{-2}\%$ & ~~~$25^{+9}_{-8}\%$\tablenotemark{c} \\
NGC~4649 & $1.4\pm0.2$ & $52^{+3}_{-4}\%$ & ~~~$47^{+6}_{-9}\%$\tablenotemark{d} \\
NGC~4365 & $2.1\pm0.6$ & $62^{+6}_{-8}\%$ & ~~~$49^{+6}_{-10}\%$\tablenotemark{a} \\
NGC~1332 & $2.2\pm0.7$ & $63^{+6}_{-9}\%$ & ~~~$30^{+11}_{-9}\%$ \tablenotemark{e} \\
NGC~1399 ($30^{\prime\prime}-90^{\prime\prime}$) & $4.7\pm1.0$ & $78^{+3}_{-4}\%$ &
~~~$68^{+8}_{-9}\%$\tablenotemark{f} \\
\enddata

\tablenotetext{a}{Sarazin et al.\ (2003)}
\tablenotetext{b}{Kundu et al.\ (2002)} 
\tablenotetext{c}{Kundu et al.\ (2003)}
\tablenotetext{d}{Randall, Sarazin, \& Irwin (2004)}
\tablenotetext{e}{Humphrey \& Buote (2004)}
\tablenotetext{f}{Angelini et al.\ (2001)}
\label{tab:compare}
\end{deluxetable}

\end{document}